# Thermal Diffusion Boron Doping of Single-Crystal Diamond

Jung-Hun Seo[1], Henry Wu[2], Solomon Mikael[1], Hongyi Mi[1], James P. Blanchard[3], Giri Venkataramanan[1], Weidong Zhou[4], Sarah Gong[5], Dane Morgan[2] and Zhenqiang Ma[1*]

[1]Department of Electrical and Computer Engineering, [2]Department of Materials Science and Engineering, [3]Department of Nuclear Engineering and Engineering Physics, [4]Department of Electrical Engineering, NanoFAB Center, University of Texas at Arlington, Arlington, TX 76019, USA, [5]Department of Biomedical Engineering and Wisconsin Institute for Discovery, University of Wisconsin-Madison, Madison, WI 53706, USA

**Abstract**: With the best overall electronic and thermal properties, single crystal diamond (SCD) is the extreme wide bandgap material that is expected to revolutionize power electronics and radio-frequency electronics in the future. However, turning SCD into useful semiconductors requires overcoming doping challenges, as conventional substitutional doping techniques, such as thermal diffusion and ion implantation, are not easily applicable to SCD. Here we report a simple and easily accessible doping strategy demonstrating that electrically activated, substitutional doping in SCD without inducing graphitization transition or lattice damage can be readily realized with thermal diffusion at relatively low temperatures by using heavily doped Si nanomembranes as a unique dopant carrying medium. Atomistic simulations elucidate a vacancy exchange boron doping mechanism that occur at the bonded interface between Si and diamond. We further demonstrate selectively doped high voltage diodes and half-wave rectifier circuits using such doped SCD. Our new doping strategy has established a reachable path toward using SCDs for future high voltage power conversion systems and for other novel diamond based electronic devices. The novel doping mechanism may find its critical use in other wide bandgap semiconductors.

With the advent of various new renewable energy sources and the emerging need to deliver and convert energy more efficiently, power electronics have received unprecedented attention in recent years. For the last several decades, Si-based power devices have played a dominant role in power conversion electronics. Wide bandgap semiconductor material based power electronics, such as those employing GaN and SiC, are expected to handle more power with higher efficiency than Si-based ones. GaN exhibits higher saturation velocity than Si. However, the thermal conductivity of GaN is low for power conversion systems. Moreover, it is currently difficult to obtain a thick and high quality GaN layer. SiC has its own native substrate, but it has inferior performance matrices (e.g., Johnson's figure of merit) versus GaN. In comparison, diamond exhibits most of the critical material properties for power electronics, except for its small substrate size at present. Diamond has a wide bandgap, high critical electric field, high carrier mobility, high carrier saturation velocities and the highest thermal conductivity among all available semiconductor materials[1-3]. Due to its superior electrical properties, the thickness of the highest quality diamond required to block an equivalent amount of voltage is approximately one-fifth to one-fourth the thicknesses of GaN or SiC. In particular, the superior thermal conductivity of diamond could greatly simplify the design of heat dissipation and hence simplify entire power electronics modules. Therefore, diamond is considered the best material candidate for power electronics in terms of power switching efficiency, reliability, and system volume and weight. However, besides the lack of large area single crystalline diamond (SCD) substrate, SCD is ultra-stable and chemically inert to most reactive reagents, due to the strong σ-bonds formed between adjacent carbon atoms, making substitutional doping very difficult[4]. Because of the doping difficulty, the majority of diamond-based diodes reported to date are Schottky diodes[5-8].

Ion implantation has been attempted to achieve substitutional doping of SCD[9,10]. However, the ion implantation process needs to be carried out at elevated temperatures (400 to 600 °C) to prevent bulk phase transition-graphitization[11-13]. In addition, a very high annealing temperature (1450 °C) under a high vacuum (~10$^{-6}$ mtorr) is required to potentially restore damaged lattices and to activate implanted dopants. During this annealing process and for high dose implantation in particular, surface graphitization still occurs, thereby creating additional unwanted processing complications for practical applications[9-11]. The alternative approach to ion implantation is in-situ doping during the epitaxial growth of diamond[2,3,14,15]. However this in-situ doping method has a number of intrinsic limitations for practical use (e.g., selective doping) due to the need for high temperature and for high density plasma during growth. For example, realizing uniform doping concentrations across a single diamond substrate using plasma enhanced epitaxial growth is rather challenging due to high plasma concentrations near the edges of diamond substrates. The small size of diamond as a substrate worsens the non-uniformity doping problem of in-situ doping. Using in-situ doping also further affects film growth uniformity and crystal quality during plasma assisted epitaxial growth. These issues make selective doping of diamond film nearly impossible, and yet selective and uniform doping is very attractive and necessary for most device applications. Previously, a solid boron film was deposited on polycrystalline diamond in order to allow boron thermal diffusion doping[7]. While a very high temperature (1600 $^{o}$C) is needed in this process and the feasibility of this method in the context of SCD doping is unknown, diodes made of polycrystalline diamond always show very high leakage current.

Here we demonstrate a very simple, yet viable method for selective boron doping in natural SCD (nSCD) via an easy thermal diffusion process at a much lower temperature than those used in any other doping methods. We use heavily doped silicon nanomembranes (SiNMs), which are now easily accessible, as our dopant carrying medium. Standalone SiNM, since its first appearance in 2005, has been extensively investigated[16]. The processing of SiNM, including transferring, doping and bonding, can be readily carried out with conventional processing tools. The SiNM is typically released in thin single-crystal silicon sheets, e.g., from a silicon-on-insulator (SOI) wafer, with thicknesses ranging from a few nanometers up to microns. The thinness of SiNM brings in a number of electronic and mechanical properties that are not possessed by its rigid counterparts (i.e., bulk Si). These unique properties have been extensively harnessed in recent years. A summary of the detailed processing methods of nanomembranes, including their creation, manipulation, etc., and their significant applications can be found elsewhere[16-18].

In this work, we first exploit the bonding advantages of SiNMs. In comparison to a regular/bulk Si substrate that is rigid and rather difficult to bond to diamond, the bonding force of SiNMs is much stronger than rigid Si due to its thin thickness. As described earlier[16], the bonding force increases exponentially with the reduction of Si thickness, which is a unique and important property of flexible SiNM. The ultra-mechanical flexibility of SiNM also allows much more conformal and higher fidelity bonding and better surface roughness tolerance when bonding with diamond, both of which facilitate intimate contacts between SiNM and diamond

(nearly 100% bonding yield without using any adhesives and no limitation on SiNM's size or shape). Also, due to the thinness of SiNM, we are able to exploit it for dopant carrying. SiNM, like bulk Si, can be heavily doped via conventional ion implantation and post-implant thermal annealing without destroying its single crystal structure. Unlike bulk Si, SiNM can be easily heavily doped across its full thickness due to its thinness. A detailed description of the method to realize SiNM full-thickness heavy doping can be found from recent publications[17-19]. Heavily boron doped thin SiNM was bonded to diamond to serve as the dopant-carrying medium for thermal diffusion in this work. As detailed in the Supplementary Information (SI, Fig. S1), the preparation and application of SiNM as a dopant carrying medium for diamond are relatively easy and straightforward. It should also be noted that heavily doped SOI wafer is also commercially available and that the doping process described here (in SI) can be considerably simplified if using such SOI wafer.

Besides the process simplicity, the SiNM based thermal diffusion doping method leads to effective, uniform and selective heavy doping in SCD. No graphitization is induced even on the diamond surface as long as it is covered by SiNM. Without involving high temperature or high vacuum furnace, the doping process can be readily realized using conventional rapid thermal annealing (RTA) process, which is widely accessible in comparison to ion implantation and in-situ doping methods. In addition, the doping method proves to be very effective in forming a shallowly doped p-type layer in diamond, which is sufficient for most device applications, while keeping its single crystallinity intact. It is noted that realizing very shallow doping in many semiconductors, such as Si, has been very challenging until the recent successful demonstration by M. L. Hoarfrost et al[56].

**RESULTS**

Fig. 1(a) illustrates the boron doping process flow. The process detail can be found in the Method section. In brief, heavily boron-doped single crystal SiNMs are first formed on an SOI substrate and then released by selective etching-away of buried oxide. The released SiNMs are transferred to diamond substrate via the stamp-assisted transfer printing method[20]. The diamond plate bearing the SiNMs is annealed via RTA to first form a stronger bonding and then to induce boron diffusion from Si into diamond. The doping mechanism is described later in the text.

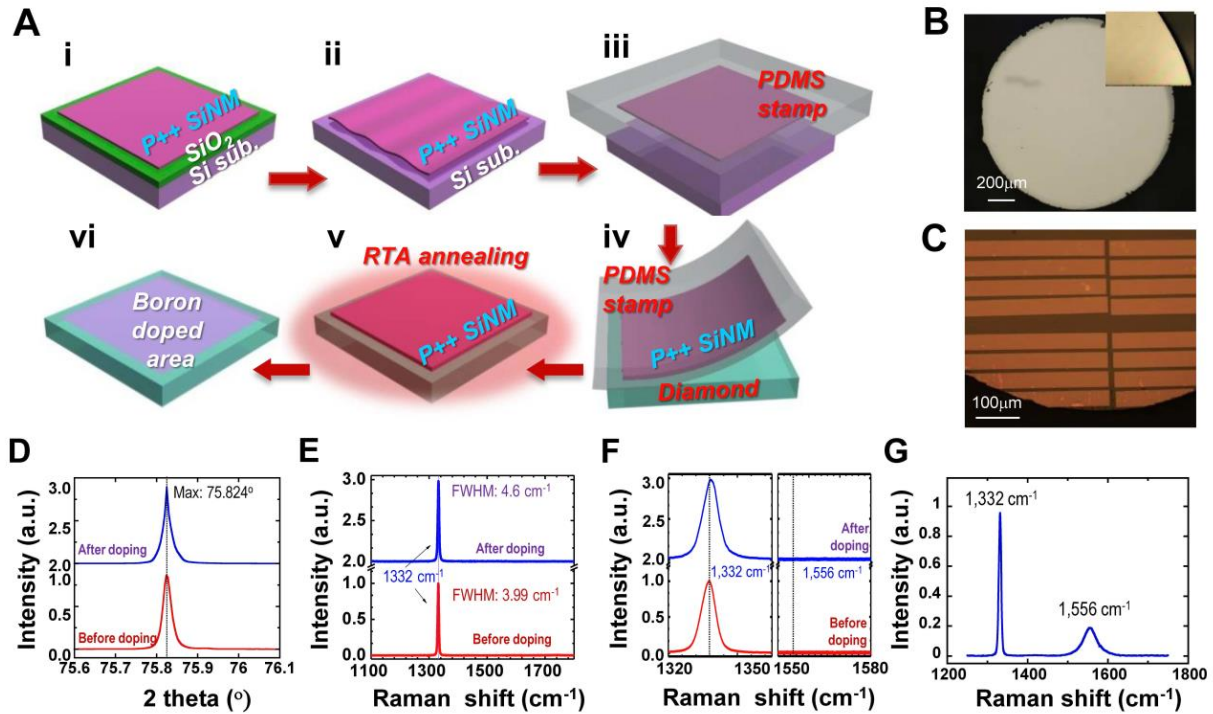

**Fig. 1. Boron doping of an nSCD plate via the thermal diffusion process.** (**a**) Process flow for boron doping in diamond plates. i. Heavy boron implantation on an SOI wafer and thermal annealing to realize heavily doped top Si on SOI. ii. Heavily boron doped top Si layer released as SiNM by selective etching of $SiO_2$. iii. Top Si picked up by an elastomeric stamp. iv. SiNM transferred to a diamond plate. v. Bond forming between SiNM and diamond and thermal diffusion with RTA. vi. SiNM removed by potassium hydroxide (KOH) etching. (**b**) Microscopic image of diamond plate before SiNM transfer. The inset shows a zoomed-in image of part of the diamond plate. (**c**) Image of diamond plate bonded with SiNM strips. (**d**) XRD spectra of the nSCD before and after boron doping. The two-theta angle is 75.82º and the FWHM is 0.025º before and after thermal diffusion, showing no noticeable change. (**e**) Raman spectra scanned from 1100 to 1800 $cm^{-1}$. The Raman shift peak is at 1332 $cm^{-1}$. (**f**) Magnified views of the Raman spectra shown in (**e**) for the ranges of 1320 -1350 $cm^{-1}$ to show the good crystallinity of nSCD and for the range of 1550 -1580 $cm^{-1}$ to show the absence of $sp^2$ bonds or distorted SiC formation after thermal diffusion. (**g**) Raman spectrum obtained from nSCD that has no SiNM bonded but is subject to the identical RTA process. The formation of $sp^2$ bonds is indicated by the Raman peak at 1556 $cm^{-1}$.

The thermal diffusion doping method has a comparative advantage over ion implantation in that lattice structural damages are not introduced during thermal diffusion. Therefore, the high temperature recrystallization process needed for post ion implantation is no longer necessary and graphitization can be readily avoided. It is also expected that higher crystal quality can be obtained using the thermal diffusion method as opposed to the ion implantation method after finishing the doping process. By using the transfer printing method, clean interfaces are ensured and importantly, selective doping (to be seen later), via deterministic transfer printing of SiNMs of different sizes to the selective areas on the diamond surface, is made easy and precise[21]. The selective doping enabled by selective transfer printing, while applicable to any size and shape of diamond plates (in contrast to direct wafer bonding), leads to a planar doped structure that can facilitate device implementations. Fig. 1(b) and (c) show images of natural diamond before and after SiNM bonding.

Diamond crystal structures before and after boron diffusion doping are first characterized and then compared. Before performing the characterizations, the SiNM is removed using potassium hydroxide (KOH) after completion of the RTA process. No graphitization removal procedures were applied to the diamond surface. After finishing the Si removal procedure, we are unable to identify any remaining Si or SiC materials in diamond from a Raman spectrum taken in the range of 400-1200 cm$^{-1}$ (see Fig. S7). The X-ray theta-2theta scan results of the boron diffused diamond are shown in Fig. 1(d). The (220) peak of the nSCD, before and after the boron diffusion process, appeared at 75.825° in both cases. Of more importance, the full width at half maximum (FWHM) of the diamond (220) diffraction peak shows no measurable changes after finishing the diffusion process. The small FWHM value of 0.025 degrees indicates the high single crystallinity of the diamond[22,23]. The Raman spectra of the nSCD before and after diffusion doping are shown in Fig. 1(e) for comparison. The zoomed-in view of the relevant wave number range of Fig. 1(e) is shown in Fig. 1(f). The $sp^3$ bonding in the sample before and after the thermal diffusion process is clearly indicated by the TO phonon peak at 1332 cm$^{-1}$. The FWHM of the Raman peak became slightly wider and shifted after diffusion (from 3.9 cm$^{-1}$ to 4.6 cm$^{-1}$ and 0.4 cm$^{-1}$ of blue-shift). Such a small change could be attributed to the change of the existing crystal imperfection in the nSCD or small stress during the process. Our nSCD before processing does not have as small FWHM value (2-3 cm$^{-1}$)[24-26] in comparison to some of the others that are reported in literature, indicating the existence of some crystal imperfection. However, the FWHM is smaller than 5 cm$^{-1}$ in both cases, further indicating that the $sp^3$ bonding in the diamond remained intact after finishing the boron diffusion process[27,28]. It is noted that the FWHM values (13.3 cm$^{-1}$ – 87.8 cm$^{-1}$)[26,29] of Raman spectra in ion implanted (after anneal) SCDs are much larger than 5 cm$^{-1}$.

The X-ray diffraction (XRD) and Raman characterizations indicate that the boron doping method via SiNM bonding and thermal diffusion does not induce measurable lattice damage in diamond. Furthermore, as shown in Fig. 1(f) it is noted that the absence of peaks near the wave numbers of 1357 cm$^{-1}$ and 1556 cm$^{-1}$ in the zoomed-in spectra of Fig. 1(e), which are the characteristic indicator of the presence of *sp$^2$* bonds, proves that the SiNM doping process does not induce detectable graphitization in the diamond bulk or on its surface. As comparison, Fig. 1(g) shows the Raman spectrum scanned from a reference nSCD sample that has no SiNM bonded but was subject to the identical RTA process. The distinct Raman peak at the wave number of 1556 cm$^{-1}$ in the spectrum clearly indicates the existence of *sp$^2$* bonds that are formed on this undoped sample. To further verify the role of SiNM in preventing graphitization on the diamond surface, the Raman spectrum taken from the backside of the SiNM bonded diamond (associated with Fig. 1(e) and 1(f)), where no SiNM is bonded, also shows a visible peak at the wave number of 1556 cm$^{-1}$ (see Fig. S3). These results indicate that it is because of using single crystal SiNM as the dopant carrying medium for thermal diffusion that we have successfully avoided graphitization on the diamond surface.

The diffused boron atom concentration in the nSCD is characterized by both secondary ion mass spectroscopy (SIMS) and capacitance-voltage (C−V) measurements. The results are shown in Fig. 2(a) and the fitted curves by the Fick's law of diffusion are shown in Fig. S4. Fig. 2(a) indicates the presence of boron at a concentration of about $1 \times 10^{19}$ cm$^{-3}$ at the diamond surface, which is comparable to the level that can be achieved by ion implantation, and gradually decreased to ~$2\times10^{15}$ cm$^{-3}$ at a depth of ~70 nm. As a comparison, ~120 cm$^2$/v·s of hole mobility and the doping concentration of ~$2\times10^{18}$ cm$^{-3}$ were measured by using Hall measurements (Accent HL5500 Hall system). Considering that the annealing time is only 40 minutes, the doping depth achieved is encouraging for device applications. The profile obtained by C−V measurements roughly matches that of SIMS in terms of shape and depth considering the limited accuracy of SIMS.

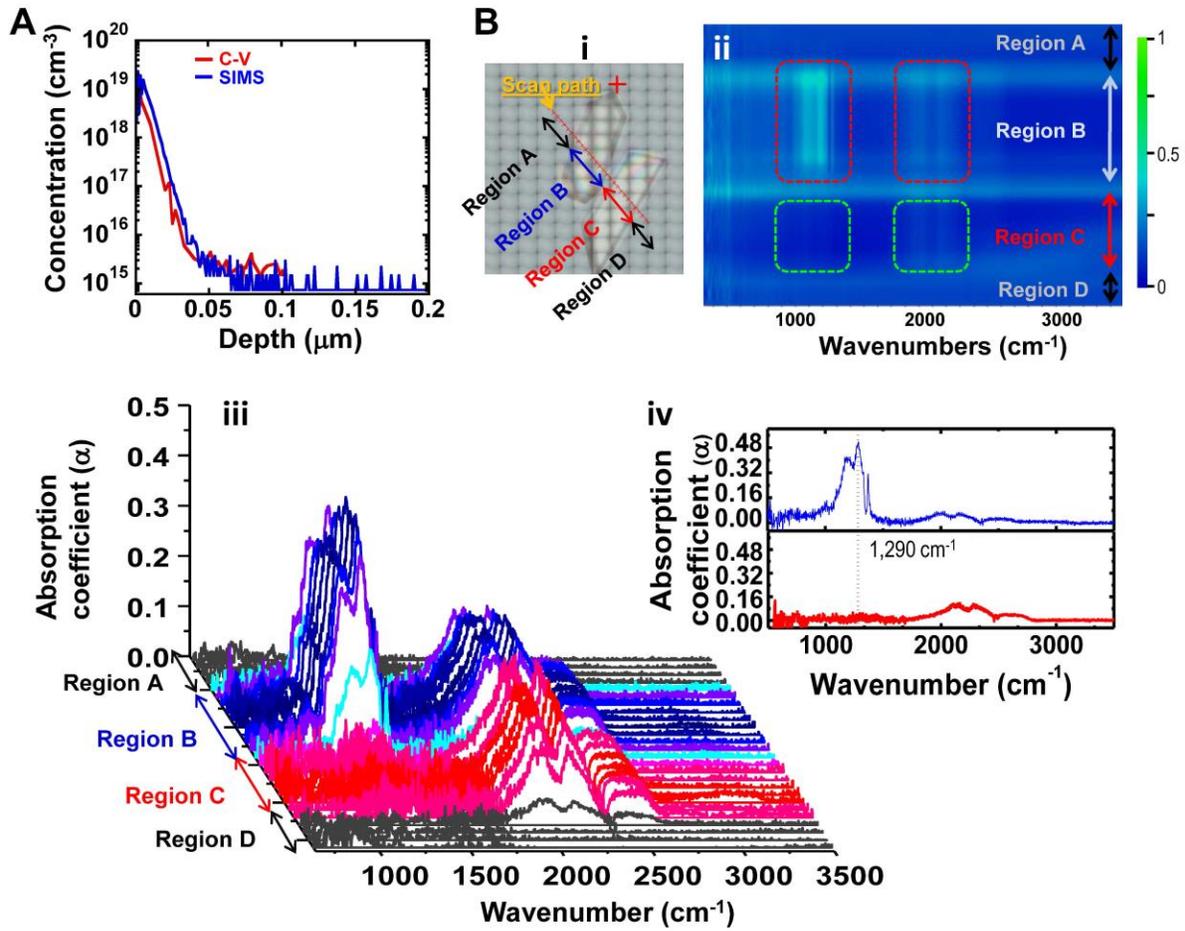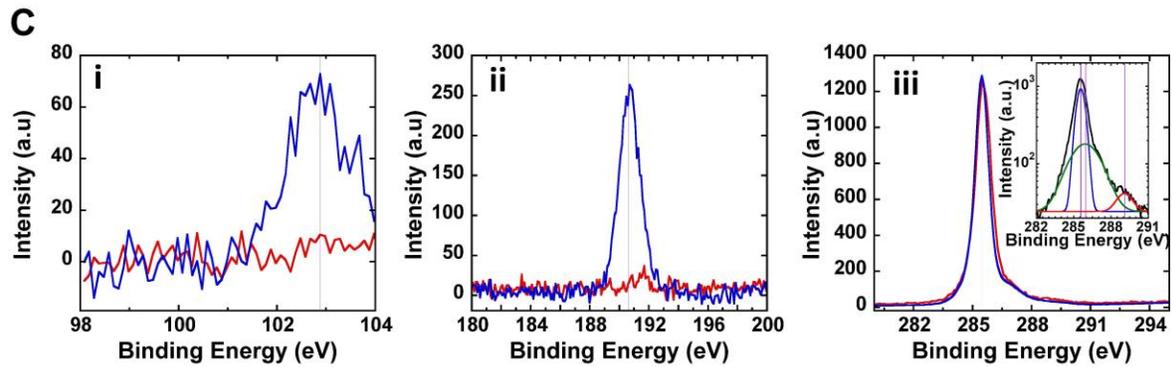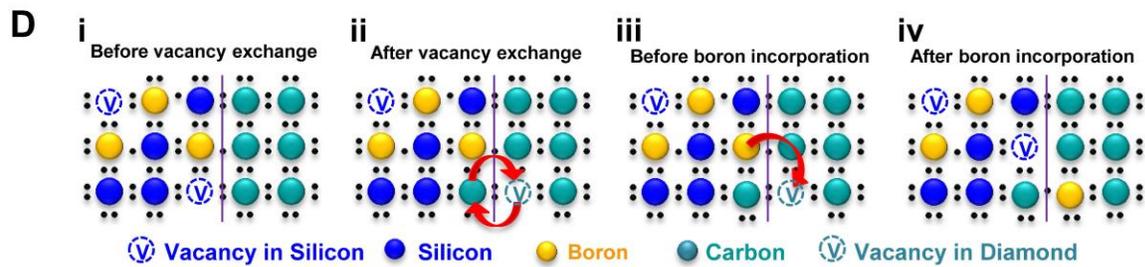

**Fig. 2. Analysis of thermally diffused nSCD. (a)** Comparison of boron doping profiles measured by SIMS and C−V methods. **(b) i.** Microscopic image of two diamond samples scanned by FTIR. The regions A and D are background. The region B is boron doped nSCD and the region C is undoped reference nSCD. **ii.** 2D FTIR scanning image corresponding to the image shown in i. The brighter color indicates stronger absorption as shown in the color scale bar. **iii.** Comparison of FTIR spectra between boron-doped nSCD and undoped nSCD. The strong absorption near the wave number of 1290 cm$^{-1}$ indicates electrically activated boron atoms; the reference sample does not have any peaks near the wave numbers of interest. **iv.** An individual FTIR spectrum obtained from regions B and C for a detailed view. **(c)** XPS spectra of boron doped (blue) and undoped (red) nSCD for **i.** Si$_{1s}$, **ii.** B$_{1s}$ and **iii.** C$_{1s}$ peaks. Inset in Fig. 2(C) iii shows zoomed-in view of the deconvoluted C$_{1s}$ peak. **(d)** Cartoon illustration of the proposed boron incorporation mechanism. **i.** Si-diamond interface just after SiNM bonding to nSCD, before thermal processing. Vacancies predominately occur on the Si-side, and all boron atoms are initially present on the Si-side. **ii.** Vacancy from the Si-side migrates across the interface, exchanging with a carbon atom from the diamond-side. This process generates a vacancy on the diamond-side at a smaller energy cost than the usual vacancy formation mechanism in diamond. **iii.** Boron atoms from the Si-side are now able to exchange with the new vacancy on the diamond-side. **iv.** Finally, boron is incorporated into diamond as a substitutional solute and can undergo accelerated diffusion by vacancies generated through steps i and ii. Please note that the schematic illustration is not meant to imply lattice matching of Si and diamond at the interface.

Further characterizations of boron-doped nSCD were performed using Fourier transform infrared spectroscopy (FTIR). Generally, boron inactivation could result from non-substitutional boron sites[30] or aggregated substitutional boron sites[31,32]. FTIR is an effective method for evaluating the substitutional doping status in a diamond. The FTIR results are shown in Fig. 2(b). Fig. 2(b) i shows two diamond plates of the same type: one is boron doped (scanning region B) that is realized using the above thermal diffusion method and the other is undoped (scanning region C) as a reference. Fig. 2(b) ii-iv shows the FTIR mapping results and the scanned spectra. The boron doped diamond shows the characteristic absorption peak at 1290 cm$^{-1}$, which clearly indicates the electrical activation of boron atoms[33]. In contrast, the peak does not appear in the undoped reference diamond. Since substitutional doping, *i.e.*, boron-carbon *sp$^3$* bonding formation is necessary for electrical activation of doped boron atoms, the FTIR and the C−V characterization results prove the substitutional doping of boron atoms in the nSCD. It should be noted that the characteristic absorption peaks associated with boron interstitials and boron interstitial complexes in diamond can be observed at 1420, 1530, 1570, and 1910 cm$^{-1}$, but no such peaks appeared in our FTIR spectra[34]. Moreover, the absence of the three infrared B-B cluster absorption peaks (553, 560, and 570 cm$^{-1}$) indicate that no aggregated substitutional boron sites were formed[35]. The broader peaks, which appear from 1900 to 2300 cm$^{-1}$ are the inherent two-phonon lines of diamond associated with C−C bonds. They appear in both the boron doped and undoped diamond samples.

X-ray photoelectron spectroscopy (XPS) was performed on both doped diamond and undoped diamond (as a reference). The binding energies for $Si_{1s}$, $B_{1s}$ and $C_{1s}$ peaks have been identified with constant pass energy of 50 eV and 100 meV energy step as shown in Fig. 2(c). The C–Si peak at ~ 103 eV indicates that a chemical reaction between Si and carbon atoms has occurred at the Si-diamond interface (Fig. 2(c) i). However, the absence of a Si peak in the Raman spectrum (Fig. S7) in the range of 400 ~ 1200 $cm^{-1}$ indicates that the C-Si reaction occurred very shallowly (< 10 nm) at the diamond surface. Boron substitution in diamond yielded a $B_{1s}$ peak at 190.6 eV corresponding to B–C (Fig. 2(c) ii) confirming that boron doping was successful. This is consistent with the SIMS, C−V and FTIR analyses. Fig. 2(c) iii shows the XPS spectra for the core $C_{1s}$ peak in the binding energy region around 280–295 eV for the undoped and boron doped nSCD samples. De-convoluted $C_{1s}$ peaks using the Gaussian/Lorentzian function in the inset of Fig. 2(c) iii shows a strong $sp^3$ C–C bonding at 285.4 eV and a very small C–O and C=O bonding at 286.4 eV and 287.9 eV, respectively, indicating that the single crystallinity of nSCD was not degraded by the boron diffusion process.

The XPS results obtained above suggest clues for elucidating the boron diffusion doping mechanisms. Especially, the C-Si bonding (Fig. 2(c) i) plays an important role for the observed boron diffusion. First-principles density functional theory (DFT) simulations[36,37] were performed to understand the diffusion mechanism. Two mechanisms were proposed to yield enhanced boron diffusion through enhanced vacancies in diamond (detailed calculation can be found in the Method section). The first mechanism involves injection of excess vacancies into the diamond from the SiNM, which has a much larger intrinsic vacancy concentration than diamond as well as excess vacancies from ion implantation into the SiNM. DFT calculations verify that this vacancy injection is much more energetically favorable than the usual mechanism for vacancy formation in diamond (movement of carbon to the diamond surface). Fig. 2(d) shows the cartoon illustration of the proposed injection and diffusion mechanism. Besides the above vacancy injection into diamond from Si, excess vacancies can also be additionally created by formation of SiC at the SiNM and diamond interface, which stabilized the vacancies by almost exactly the required 0.7 eV needed to explain the enhanced boron diffusion rate into diamond. Both mechanisms could play a role and further research is required to elucidate their contributions. In each case, diffused boron atoms from Si are expected to immediately become substitutional atoms in diamond if the diamond does not have pre-existing vacancy related defects (ideal situation). Since no vacancy defects are additionally generated in diamond during the diffusion process, a high temperature anneal that is necessary for post-implantation[38] then becomes unnecessary in such a doping process.

Based on the identified vacancy exchange mechanism, it is unlikely that boron atoms diffuse to non-substitutional sites, which is different from ion implantation induced case[30,39]. It is also noted that the realized boron doping concentration in our method is much lower than that realized in synthetic diamond[31,32] and therefore the likelihood of forming aggregated boron substitutional sites should be rather small.

Substitutional boron doping in diamond using thermal diffusion has never been considered possible in the past. Our experiment clearly demonstrated the viability of boron doping through the thermal diffusion process using heavily doped, bonded SiNM as dopant carrying medium. The process is also simple and easily accessible. We propose that the origin of the successful doping is enhanced boron diffusion into diamond enabled by the Si-C bonding states near the Si-diamond interface, which lower the energy to create diamond vacancies and enhance boron transport.

The graphitization-free process is thought to be directly related to the intimate bonding between Si and diamond. A previous study has shown that introducing impurities at elevated temperature in diamond can prohibit graphitization during thermal annealing[13]. In the current thermal diffusion experiment, boron atoms accumulate at the bonded surface of the diamond as soon as thermal diffusion begins to occur at an elevated temperature. Therefore, no phase transition is expected to happen at the SiNM bonded diamond surface under the special thermal diffusion setting. This expectation is consistent with experimental observations.

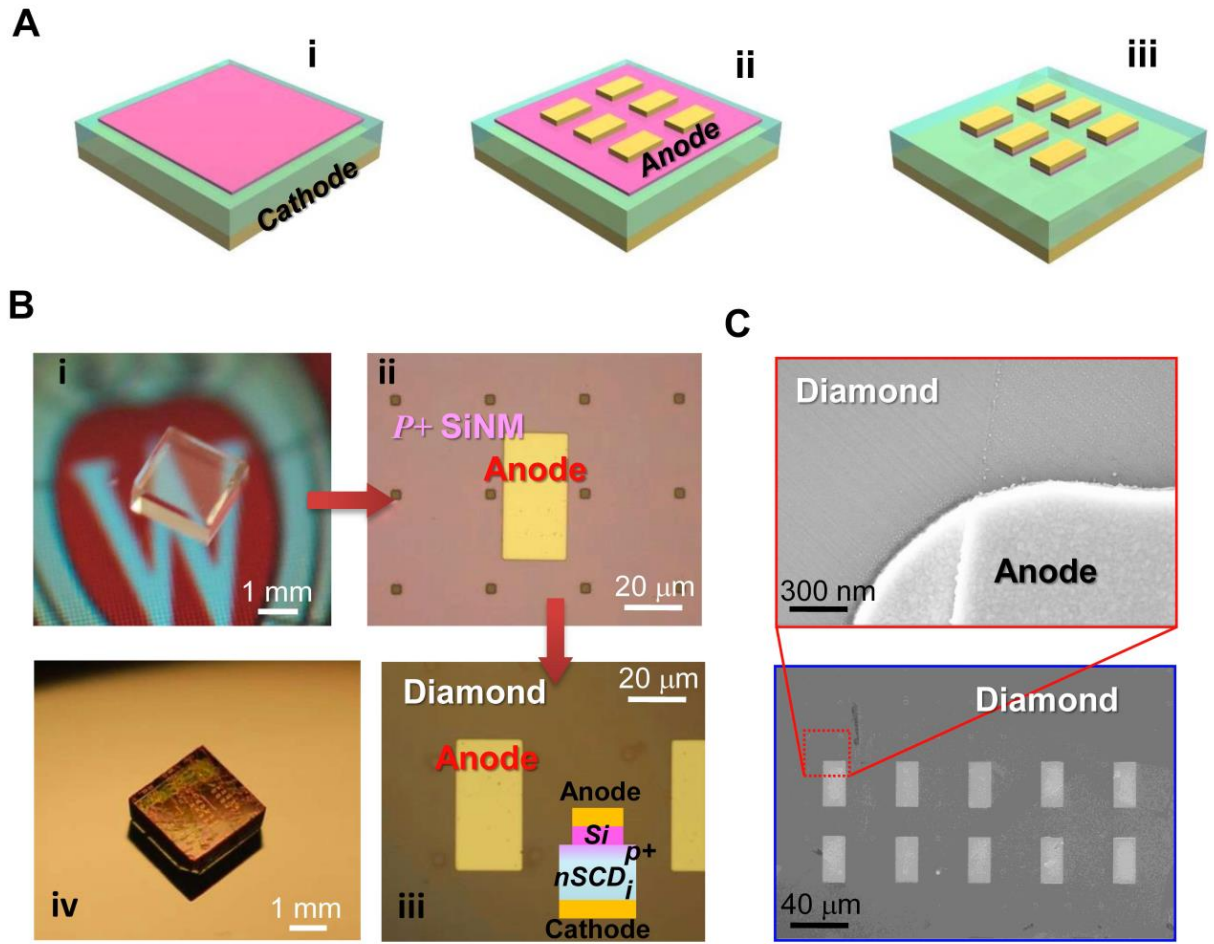

**Fig. 3. Fabrication of nSCD diodes.** (**a**) Process illustration of *p-i* nSCD diode fabrication after completion of the thermal diffusion doping processing steps (after the step shown in Fig. 1(A) v). **i.** Form backside cathode metal. **ii.** Deposit anode metal on top of SiNM that is bonded to nSCD. **iii.** Etch SiNM and diamond using anode metal as etching mask. (**b**) **i.** Optical image of nSCD before processing. **ii.** Microscopic image corresponding to the processing step shown in (A) ii. **iii.** Microscopic image of finished device corresponding to step shown in (A) iii. Each diode's area is 800 μm². The inset shows the cross section illustration of the diodes. **iv.** Image of a diode array on an nSCD plate. (**c**) SEM and zoomed-in images of finished diodes.

The above boron doping method is used to fabricate diodes using a 2 × 2 mm², 120 μm thick nSCD plate (Fig. 3). Fig. 3(a) shows the process flow for fabricating vertical *p-i* junction diodes. After completion of SiNM bonding and boron thermal diffusion from SiNM to diamond (Fig. 1 (a) v) the cathode is formed first on the bottom side of the diamond plate. The anode is formed directly on top of the SiNMs (Fig. 1 (a) ii), which previously served as a boron carrying medium, since it is much easier to form ohmic contacts on heavily p-type doped Si. It is noted that a *p-i* junction is not formed between Si and diamond in this case, but between the p-type doped diamond surface and the intrinsic diamond bulk. Both the anode and the cathode contacts are ohmic contacts. To prove that the *p-i* junction is formed in diamond not in between Si and diamond, a *p-i* diamond diode without an SiNM layer, which was removed after completion of boron diffusion, was fabricated. The results are shown in the SI (Fig. S6). Fig. 3(b) shows (i) the optical images of the diamond before processing, (ii) the SiNM bonded diamond, (iii) finished diodes, and (iv) the diamond diode array. Fig. 3(c) further shows the scanning electron microscope (SEM) images of the finished diamond diodes. As can be seen here, using the transfer printed and patched SiNMs, selective doping can be easily realized on the diamond surface.

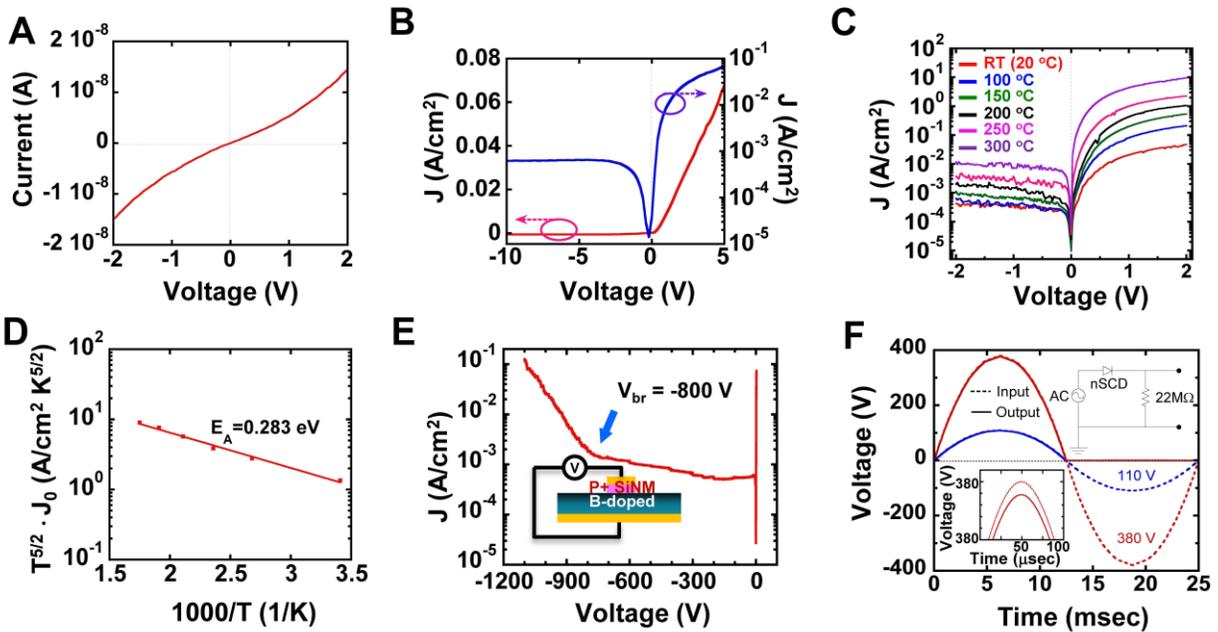

**Fig. 4. Electrical characteristics of nSCD diodes.** (**a**) Measured I-V characteristics from two adjacent bottom metal contacts, indicating ohmic contact behavior. (**b**) Measured I-V characteristic of vertical junction nSCD diode within a low bias range (-10 V ~ 5 V). (**c**) Measured I-V characteristics of diodes under different temperatures. (**d**) Arrhenius-fit plot: $T^{5/2} \cdot J_0$ versus inverse temperature (room temperature to 350 °C) at zero bias for extraction of boron activation energy. The data fitting gives a boron activation energy of 0.283 eV. (**e**) Measured I-V characteristic of vertical junction nSCD diodes within a high bias range. Breakdown occurs at around -800 V. (**f**) Characteristics of a half-wave rectifying circuit for 60 Hz AC voltage input in one period. The bottom inset shows a zoomed-in view of a 380 V curve for 12.5 msec. The top inset shows the circuit schematic with the nSCD diode connected with a 0.5 W 22 MΩ load.

Fig. 4(a) shows the current-voltage (I−V) characteristics from two adjacent bottom contacts formed on the diamond surface. The nearly straight I−V curve indicates that acceptable ohmic contacts are formed. Fig. 4(b) shows the measured forward and reverse bias (up to -10 V) I−V characteristics of the *p-i* nSCD diode. The diode shows good rectifying behavior and the ideality factor is found to be 1.3. The high ideality factor is ascribed to surface conductive channels which are induced during the doping process. The current density in the vertical junction diode is 0.07 A/cm$^2$ at +5 V, which is lower in comparison to diamond Schottky diodes[5-8]. The low current density is mainly due to the low carrier concentration in the undoped nSCD (see SI for more comparison analysis). Fig. 4(c) shows the measured diode's I−V characteristics at various temperatures (from room temperature to 300 °C) in order to extract the boron activation energy. As the temperature increased, both forward and reverse currents increased, with the forward current increasing faster than the reverse current, leading to an improved rectification ratio and forward ideality factor. The temperature-dependent current behavior is resulted from desirable semiconducting properties of the diamond, similar to the Schottky diodes reported earlier[6], which further indicates the potential high temperature application of the diodes. Fig. 4(d) shows the Arrhenius plot of the diodes under zero bias, which is used to obtain the actual activation energy of the boron atoms. Using $J_0 \propto T^{5/2} \exp(-E_a/kT)$, where $E_a$ is the activation energy[40], we obtain 0.283 eV for $E_a$, which agrees well with reported values[51] under boron concentration reported in this work.

Fig. 4(e) shows the large reverse bias range measurement results of the diodes. The reverse bias current increases less than one order from 0 V to -800 V, where diode breakdown begins to appear. To our knowledge, this is the highest breakdown voltage that has been reported from a diamond diode. It is also the first *p-i* diamond diode realized using non in-situ doping methods[3]. Considering the shallow boron doping depth that is achieved here, the results indicate the great potential of SCD for power rectification and switching. High quality synthetic diamond has a breakdown electric field of about 10$^7$ V/cm. The breakdown electric field of natural diamond is usually about a quarter to half of the value of synthetic diamond. The diamond plate used in this study is 120 μm. A simple calculation reveals that the nearly intrinsic bulk diamond is far from being fully depleted before diode breakdown occurs. Since the boron doping depth in the diamond is very shallow and the doping concentration quickly degrades from the top surface (Fig.

2(a)), the boron doped top layer has been fully depleted under certain reverse bias. Hence, further depletion of the bulk diamond is impossible, thus limiting the breakdown voltage of the diodes. It is expected that a higher breakdown voltage would be achieved if the boron doping depth and concentration could be improved (note: 30 to 60 kV breakdown voltage is estimated under full depletion of the 120 μm bulk of this nSCD).

A half-wave rectifier circuit was implemented using one of the fabricated *p-i* diodes. Fig. 4(f) shows the input and output waveforms of the circuit, which is shown in the inset, using AC voltages of 110 V and 380 V, respectively. Due to the excellent block voltage and low turn-on voltage of the diodes, the circuit exhibits very good rectifying characteristics. Using a 22 MΩ load resistor, the forward voltage drop is 2.2 V and 1.4 V for 380 V and 110 V ac voltages, respectively. Since the bulk region of the diamond diode is not fully depleted under reverse bias, the high series resistance (~1.1 kΩ) of the diode contributed significantly to the forward voltage drop of the rectifier.

**DISCUSSION**

Heavily boron doped single crystal Si nanomembranes serve as unique boron dopant carriers for practically doping single crystal diamonds. The intimate bonding between silicon and diamond supports a novel vacancy exchange-based boron doping mechanism. The new doping mechanism dramatically lowers the thermal diffusion temperature that is needed to realize substitutional boron doping, without inducing any graphitization or lattice damage, in single crystal diamonds. The selective boron doping enabled by transfer printed Si nanomembrane lead to the demonstration of *p-i* diamond diode arrays with excellent rectifying and voltage blocking characteristics. This lattice-damage-free and phase-transition-free doping method, as well as its easy application process may find critical use not only in n-type diffusion doping into SCD but also in other wide bandgap semiconductors.

## Methods

**Device Fabrication and Characterizations.** Commercially available chemical mechanical polished type-IIa nSCD (from Harris International) plates are used in this study. The different pieces of nSCDs used in this study are characterized to have the same properties. It should be noted that the method also worked with CVD SCD and the relevant results with CVD SCD are under preparation. For SiNM bonding to diamond, the fabrication began with an SOI wafer (from Soitec) with a boron doped 200 nm top Si layer and a doping level of $4\times10^{15}$ cm$^{-3}$. A 30 nm of thermally grown SiO$_2$ on an SOI wafer was used as a screen oxide layer prior to ion implantation. Ion implantation was carried out with boron ions at an energy level of 16 KeV and a dose of $3\times10^{15}$ atoms/cm$^2$ at a 7 degree incident angle. Furnace annealing then takes place at 950 $^{o}$C for 90 min under nitrogen ambient. During annealing, Si recrystallization and dopant redistribution occurred. After dopant redistribution, boron concentration at the bottom side of the top Si layer reaches ~$10^{20}$ cm$^{-3}$ (see Fig. S1). After photolithography and reactive ion etching (RIE, Unaxis 790) steps for defining the etching holes on the top Si layer are completed, the 145 nm buried oxide layer of the SOI was undercut with concentrated hydrofluoric acid (HF, 49%). The released top Si layer, becoming SiNM, was transfer printed onto nSCD plates. Detailed transfer printing procedures can be found elsewhere[15,17,19]. Prior to the transfer printing step, the nSCD plate was immersed in an ammonium sulphuric acid solution for 30 min at 200 $^{o}$C and then rinsed in an ammonium hydroxide/hydrogen peroxide solution, followed by a deionized (DI) water rinse to obtain a contaminant-free and native oxide-free surface. The nSCD plates with transferred boron doped SiNM on top were annealed in an RTA for 40 min at 800 $^{o}$C under nitrogen ambient.

With regard to diode fabrication, after finishing thermal diffusion (the step shown in Fig. 1(a) v) the cathode electrode formed on the bottom side of the nSCD plate, as shown in Fig. 3(a) i, with Ti/Pt/Au (20/50/100 nm) by e-beam evaporation, followed by annealing at 450 $^{o}$C with RTA. Next, the anode metal of Ti/Au (20 nm/150 nm) formed on top of the SiNM. For diodes without a Si layer (Fig. S6), the top Si layer was removed using dry etching before forming the top anode metal contacts (Ti/Pt/Au: 20/50/100 nm). The metal pad size formed on both diamond and SiNM is 20 μm × 40 μm with 35 μm distance in between the metal pads. Using the anode metal pads on SiNM as an etching mask, the SiNM around the metal pads was dry etched away till the diamond surface is exposed (Fig. 3(a) iii). Further etching by RIE with oxygen was carried out to etch the diamond surface 50 nm down from the initial diamond surface to minimize the current flow on the diamond surface. I−V and C−V characteristics were obtained using an Agilent 4155B semiconductor parameter analyzer, Keithley 237 high-voltage source-measure unit and an Agilent E4980A precision LCR meter in the dark to avoid any light induced photocurrents.

XPS was performed using a Thermo K-alpha spectrometer that is equipped with a monochromated Al Kα X-ray source. The vacuum of the spectrometer during measurements was less than $2\times10^{-10}$ Torr.

**Calculation for boron diffusion from SiNM into Diamond.** From Fig. 2(a) we fit the concentration profile to Fick's law of diffusion from a constant source and find the effective diffusion coefficient of boron in diamond near the diamond SiNM interface at 800 ℃ to be $2\times10^{-16}$ cm$^2$/s. This diffusivity is approximately three orders of magnitude higher at 800 ℃ than previous implantation experimental measurements of boron diffusion in diamond, which have also found an activation barrier of approximately 4.2 eV[42]. The enhanced diffusion observed in the present study at 800 ℃ would correspond to an activation barrier decrease of approximately 0.7 eV, assuming the overall diffusion constant $D_0$ prefactor remains unchanged. To understand the increased boron transport into diamond from SiNM, the atomistic mechanism of boron migration needs to be determined. From first-principles density functional theory (DFT) simulations, the formation energy of interstitial boron in diamond is found to be 10 eV higher than that for substitutional boron[32]. This suggests that boron diffusion in diamond likely occurs through a vacancy-mediated mechanism. Under this mechanism, both the vacancy formation energy and the vacancy migration energy would contribute to the diffusion barrier of boron.

In this study we calculate the neutral diamond vacancy formation energy to be 6.29 eV with DFT, where this value includes the estimated 0.36 eV Jahn-Teller relaxation energy for the diamond vacancy[54]. This formation energy is within previously calculated range of 6-7 eV[43,44], though the formation energy lowers significantly for positively charged vacancies in p-type diamond[43]. The vacancy migration energy for the neutral diamond vacancy has been experimentally measured to be 2.3 eV[45], with DFT results ranging from 1.7-2.8 eV[43,44]. We find the DFT barrier for boron-vacancy exchange in diamond to be 1.47 eV, showing that boron diffusion is limited by bulk vacancy migration.

We propose that the observed increase in boron diffusion results from increased diamond vacancy formation at the Si-diamond interface. During annealing, interactions can occur between Si and C at the Si-diamond interface, and the energy gained from these reactions can help lower the formation energy of diamond vacancies. We consider two possible interactions with Si at the interface, vacancy injection and SiC formation.

Due to the much larger intrinsic vacancy concentration on the silicon side, as well as excess vacancies from ion implantation into the SiNM, we expect that carbon atoms can jump into Si vacancies. This essentially results in an "injection" of vacancies from the SiNM into the diamond.

An estimate of the reaction energy for vacancy injection of an existing vacancy from the SiNM into the diamond can be calculated by the following equation:

$$E_{Form} = Products - Reactants = (E_{Diamond}^{Vac} + E_{Si}^{C\text{-}Sub}) - (E_{Diamond}^{Bulk} + E_{Si}^{Vac}) \quad [1]$$

where $E_{Diamond}^{Bulk}$ is the energy of a bulk diamond supercell, $E_{Diamond}^{Vac}$ and $E_{Si}^{Vac}$ are the energies of a single vacancy within a diamond supercell and Si supercell, respectively, and $E_{Si}^{C-Sub}$ is the energy of a Si supercell with a single C substitution. $E_{Form}$ is the energy of the injection reaction, which is the energy needed to form a diamond vacancy and a carbon substitution in Si, starting from bulk diamond and a Si vacancy. From DFT calculations, we obtain 4.12 eV as the energy of this reaction, much less than the 6.29 eV needed to form a vacancy in pure diamond calculated with corresponding methods and approximations. Thus injection enables the formation of more diamond vacancies that are available under normal equilibrium conditions. Note that the 2.17 eV reduction in the vacancy formation energy for diamond is only available when there are excess vacancies available in the SiNM that can be accessed without any energy cost. Once these excess vacancies are gone continuing this reaction would require also contributing the energy to form a Si vacancy, at which point the reaction becomes energetically unfavorable compared to the usual diamond vacancy formation mechanism.

If the reaction at the Si-diamond interface produces SiC rather than a substitutional C in Si, then the energy gained from the reaction can be written

$$E_{Form} = Products - Reactants = (E_{Diamond}^{Vac} + E_{SiC}) - (E_{Diamond}^{Bulk} + E_{Si}^{Atom}) \quad [2]$$

where $E_{SiC}$ is the energy of a pair of SiC atoms, $E_{Si}^{Atom}$ is the energy of a single Si bulk atom, and other quantities are as previously defined. The SiC heat of formation, from Si and diamond, has been determined experimentally to be -0.68 eV per formula-unit (one SiC atom pair)[55]. Thus $E_{Form}$ from reaction (2) gives 5.61 eV, which lowers the vacancy formation energy by almost exactly the required 0.7 eV needed to explain the enhanced boron diffusion. Please note that the amount of vacancy formation energy reduction is solely due to the experimental SiC heat of formation and is not dependent on any specific DFT calculation. This agreement is suggestive, although it may simply be coincidence. Furthermore, although no SiC is found in the Raman characterization of the doped sample (but shown in the XPS results), it indicates that only a small number of vacancies is needed to provide the enhanced diffusion. Overall, more study is necessary to determine which mechanism is dominant.

**Computational Methods.** We performed first-principles density functional theory (DFT) calculations with the Vienna Ab-initio Simulation Package (VASP)[46-49]. Exchange–correlation was treated in the Generalized Gradient Approximation (GGA), as parameterized by Perdew, Burke, and Ernzerhof (PBE)[50,51]. The projector augmented wave method (PAW)[52,53] pseudopotentials utilized the following valence configurations: $3s^2\ 3p^2$ for Si, $2s^2\ 2p^2$ for C, and $2s^2\ 2p^1$ for B. Bulk and defect calculations were done using 2×2×2 cubic supercells of Si and C, with 64 atoms. The energy cutoff was set to 900 eV, while the Brillouin zone was sampled by a 4×4×4 Monkhorst–Pack k-point mesh. Errors in energy are converged to less than 1 meV/atom with respect to the energy cutoff and k-points.

## Acknowledgments

This work was supported by an Air Force of Scientific Research (AFOSR) under a Presidential Early Career Award for Scientists and Engineers (PECASE) # FA9550-09-1-0482. The program manager at AFOSR is Dr. Gernot Pomrenke. H.W., D.M. and the molecular modeling work were supported by the National Science Foundation Software Infrastructure for Sustained Innovation (SI2), award No. 1148011. S. M. acknowledges the support by Graduate Engineering Research Scholars (GERS) fellowship and Winslow Sargeant fellowship.

Supplementary Materials for

# Thermal Diffusion Boron Doping of Single-Crystal Diamond

Jung-Hun Seo[1], Henry Wu[2], Solomon Mikael[1], Hongyi Mi[1], James P. Blanchard[3], Giri Venkataramanan[1], Weidong Zhou[4], Sarah Gong[5], Dane Morgan[2] and Zhenqiang Ma[1*]

[1]*Department of Electrical and Computer Engineering,* [2]*Department of Materials Science and Engineering,* [3]*Department of Nuclear Engineering and Engineering Physics,* [4]*Department of Electrical Engineering, NanoFAB Center, University of Texas at Arlington, Arlington, TX 76019, USA,* [5]*Department of Biomedical Engineering and Wisconsin Institute for Discovery, University of Wisconsin-Madison, Madison, WI 53706, USA*

[*]Authors to whom correspondence should be addressed. Electronic address: mazq@engr.wisc.edu

## 1. Boron doping profile in SOI and removal of SiNM after finishing thermal diffusion:

Fig. S1 shows the boron doping profile of SOI before and after annealing. Heavily boron doped Si ($>10^{20}$ cm$^{-3}$) can be removed by KOH, although the etching rate will be slower than undoped and lightly doped Si[1]. The SiNM is removed with KOH after finishing the thermal diffusion process.

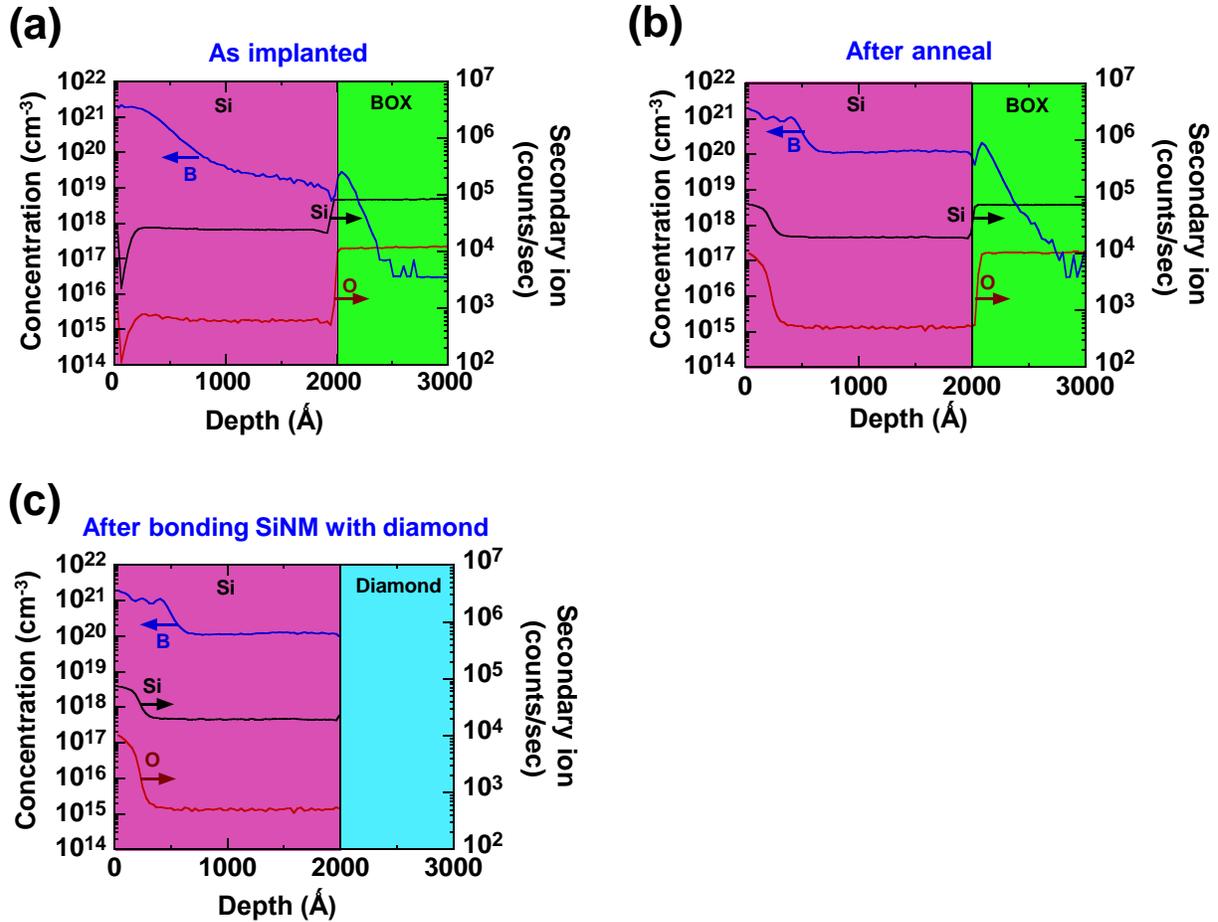

**Fig. S1.** (a) SIMS profile of SOI after completion of ion implantation. (b) SIMS profile after completion of thermal annealing. The boron concentration near the bottom of the top Si layer in the SOI reaches $10^{20}$ cm$^{-3}$. (c) Illustration the situation when SiNM is bonded to diamond, before RTA is applied.

## 2. Natural single crystalline diamond (nSCD) plates:

The starting diamond plates have polished surfaces. As shown in Fig. S2(a), scanning electron microscope (SEM) images show tiny grooves on the diamond surface, which result from the surface polishing step. In Fig. S2(b), the atomic force microscopy (AFM) surface scanning results confirm that the surface roughness of the direction perpendicular to the groove direction is 1.05 nm, while the roughness parallel to the groove is 0.35 nm. The surface roughness difference along different directions is attributed to the mechanical polishing of diamond[2]. Under such surface smoothness conditions, transfer SiNM to a diamond plate without using adhesive is of no difference in applying the SiNM transfer techniques to a silicon substrate.

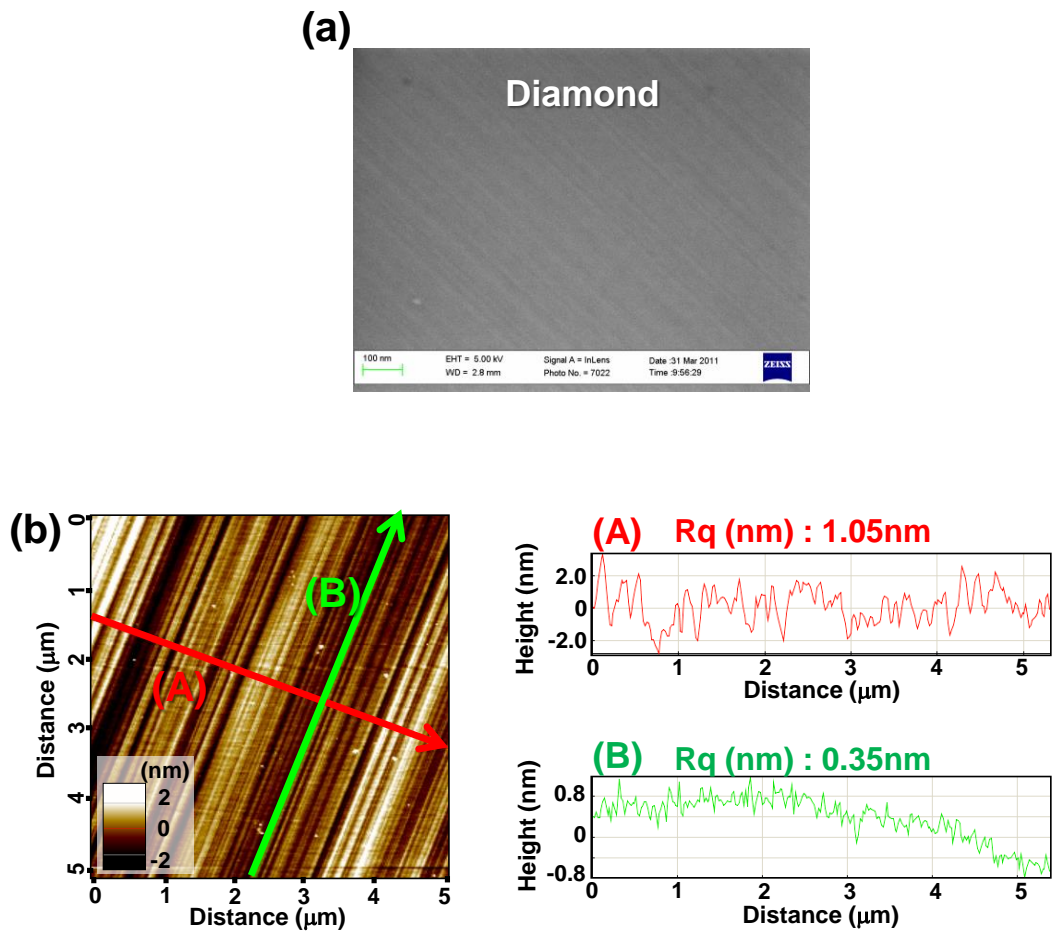

**Fig. S2. (a)** Planar SEM image of polished nSCD surface. **(b)** AFM image of nSCD before SiNM bonding and surface roughness shown by AFM scanning along two perpendicular directions.

### 3. XRD, Raman, SIMS and FTIR characterizations:

XRD was performed using a PANalytical X'Pert PRO diffractometer with Cu Kα radiation. Raman analysis was carried out using a Horiba LabRAM ARAMIS Raman confocal microscope with 18.5 mW of He-Ne (632.8 nm) laser light. The spectrometer resolution is 0.045 cm$^{-1}$. The SiNMs that were bonded to diamond were removed using KOH before performing the XRD and Raman analyses. A SIMS profile was obtained from QSPEC Technology, Inc. FTIR spectra were acquired using a NICOLET iN-10 FT-IR (Thermo Scientific) with a spectral resolution of 2 cm$^{-1}$ in the scanning range from 500 cm$^{-1}$ to 3500 cm$^{-1}$.

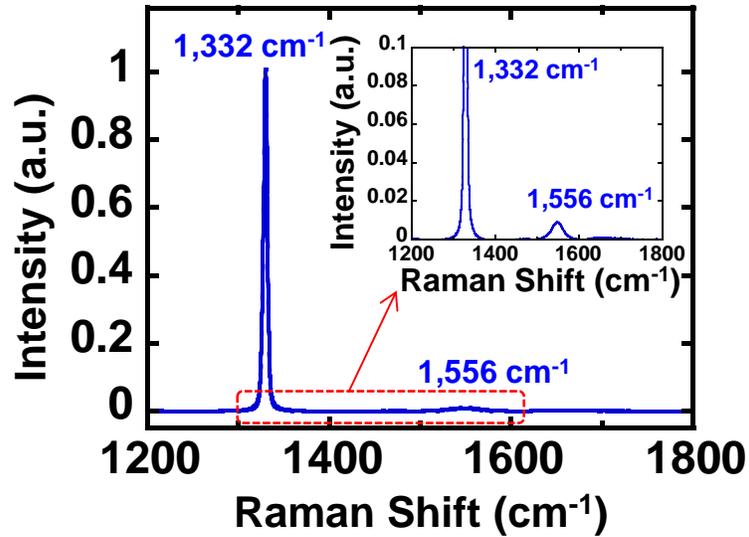

**Fig. S3.** Raman spectrum obtained from the backside of nSCD. The front side of the nSCD is bonded with SiNM and the sample was subject to RTA. The visible peak at the wave number of 1556 cm$^{-1}$ indicates the formation of $sp^2$ bonding, which is in contrast to the Raman spectrum obtained from its front side.

## 4. Fitting boron profile by the Fick's law of diffusion

Each of measured boron doping profiles measured by SIMS and C−V methods were fitted by the equation for Fick's laws of diffusion from constant source:

$$(C(x,t) - C_{bulk}) / (C_{source} - C_{bulk}) = erfc\left(\frac{x}{2 \cdot \sqrt{D \cdot t}}\right)$$

where, $C_{bulk}$ is the initial concentration in the bulk, ~ $2 \times 10^{-15}$ cm$^{-3}$, which is the lowest detection value in C−V and SIMS methods and $C_{source}$ is the source concentration at $x=0$ and $t$ is the time, in this case it's 40 minutes (2400 seconds).

For the SIMS fit, $C_{source}$ is approximately $3 \times 10^{19}$ cm$^{-3}$, D is approximately $1.6 \times 10^{-16}$ cm$^2$/s.
For the C−V fit, $C_{source}$ is approximately $1 \times 10^{19}$ cm$^{-3}$, D is approximately $1.6 \times 10^{-16}$ cm$^2$/s.

Under such conditions, measured boron doping profiles match well with the fitted result (orange colored curve).

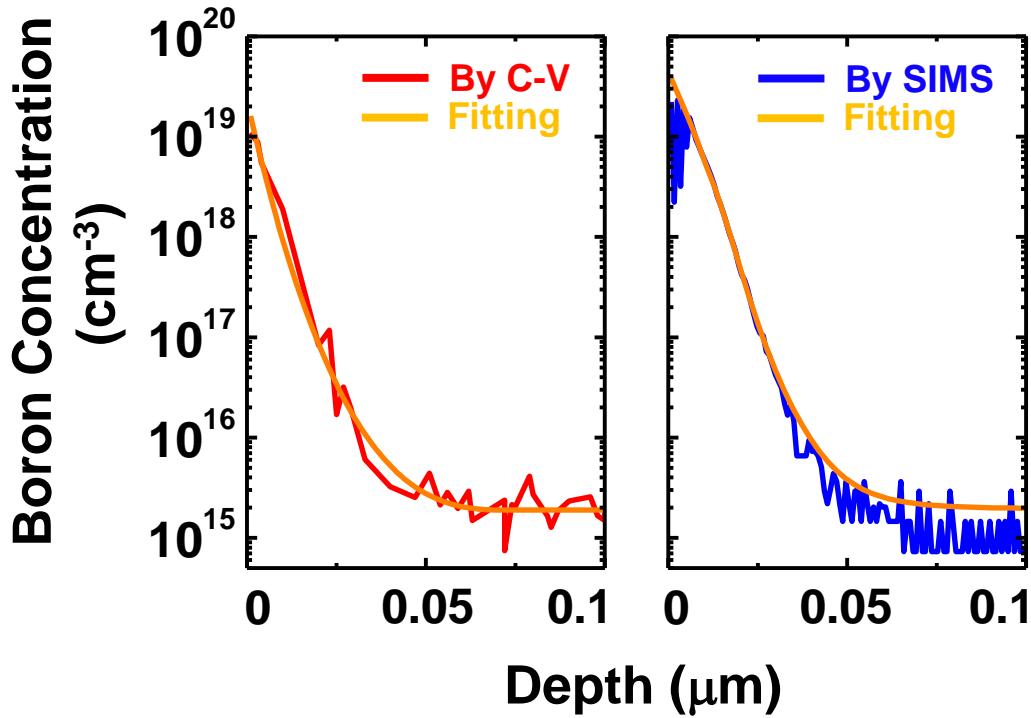

**Fig. S4.** The Measured and fitted boron profiles on nSCD samples.

## 5. C-V measurements to characterize boron doping profile and corresponding band diagram:

The depletion-layer capacitance per unit area is given by:

$$C = \frac{\varepsilon_s}{W_D} = \sqrt{\frac{q \cdot \varepsilon_s \cdot N}{2}} \left( \psi_B - V - \frac{2kT}{q} \right)^{-1/2}$$

The $1/C^2$ -V relationship is obtained by rearranging the above equation:

$$\frac{1}{C^2} = \frac{2}{q \cdot \varepsilon_s \cdot N} \left( \psi_B - V - \frac{2kT}{q} \right)$$

Then the carrier doping concentration can be extracted by taking the derivative of $1/C^2$ with respect to voltage:

$$\frac{d(1/C^2)}{dV} = -\frac{2}{q \cdot \varepsilon_s \cdot N}$$

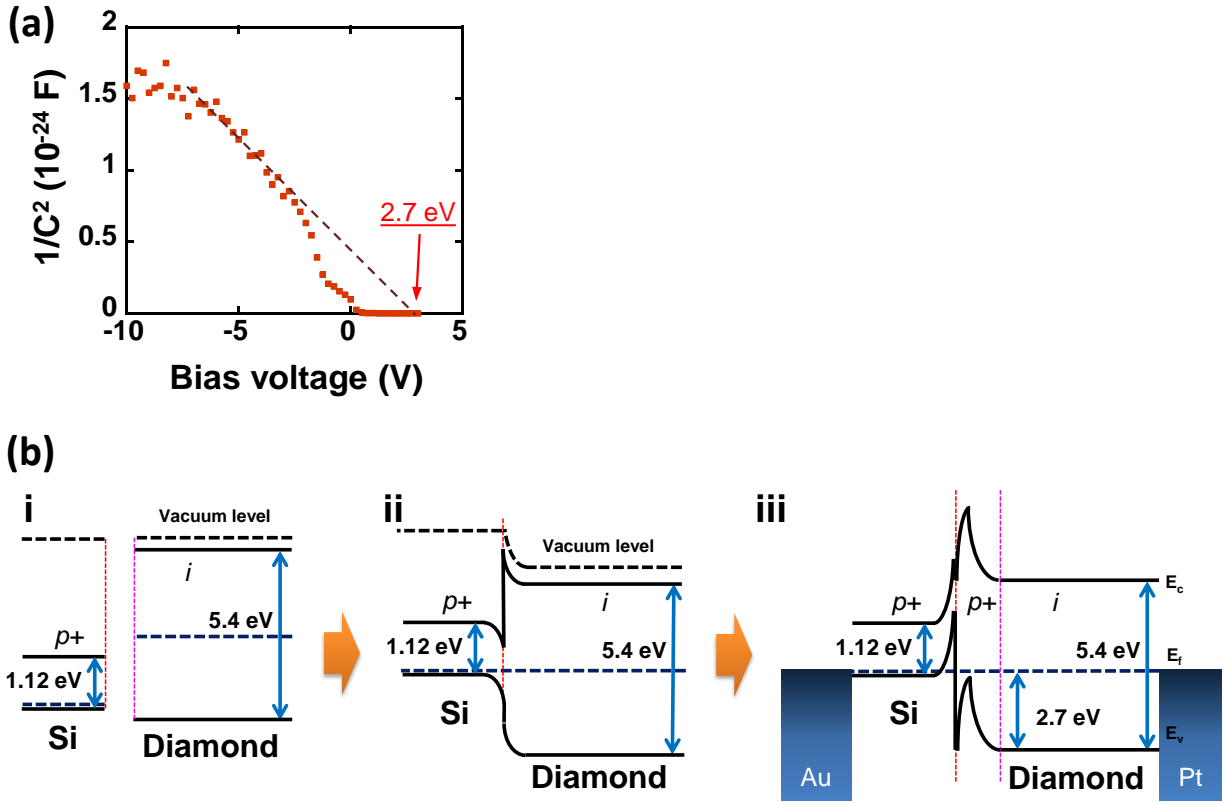

**Fig. S5.** (a) The $1/C^2$ versus voltage plot measured from derivative -10 V to 5V. It is used to extract the boron doping concentration and built-in potential of the *p-i* junction. (b) Band

diagrams. (i) Band alignment before bonding *p*-Si with *i*-nSCD. (ii) Band diagram after bonding *p*-Si and *i*-nSCD before anneal/boron diffusion. (iii) Band diagram after completion of thermal anneal/boron diffusion.

## 6. I-V characteristic of *p-i* nSCD diode without Si NM

Fig. S6 shows the measured I−V characteristics of a *p-i* nSCD diode where SiNM was removed after completion of the thermal diffusion process.

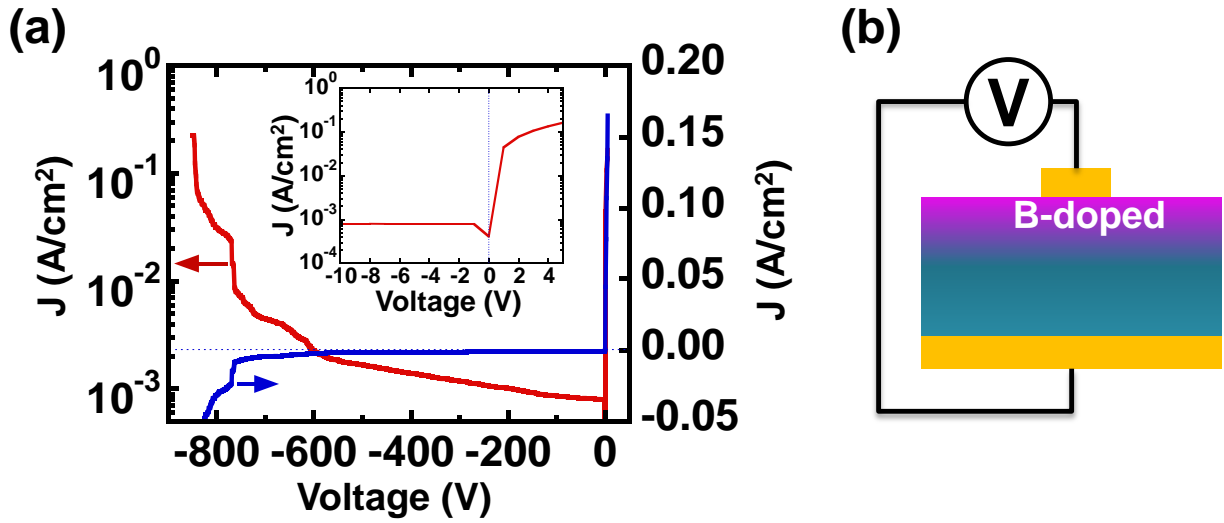

**Fig. S6. (a)** Measured I−V characteristics of a vertical junction nSCD diode (without SiNM in the diode's vertical structure) in a bias range from -850V to 10V. **(b)** Illustration of the diamond diode structure.

## 7. Raman spectroscopy analysis to check the existence of Si and SiC

Fig. S7 shows the Raman spectra in the range of 400 ~ 1200 cm$^{-1}$ that are taken from an undoped nSCD and from a diffused nSCD after the finishing removal of SiNM. The two spectra are compared. The comparison shows no difference in terms of characteristic peaks. As shown in Fig. S7 (b), the doped nSCD surface doesn't show any characteristic peaks of Si or SiC, which otherwise should appear at 520 cm$^{-1}$ and around 800~900 cm$^{-1}$, respectively.

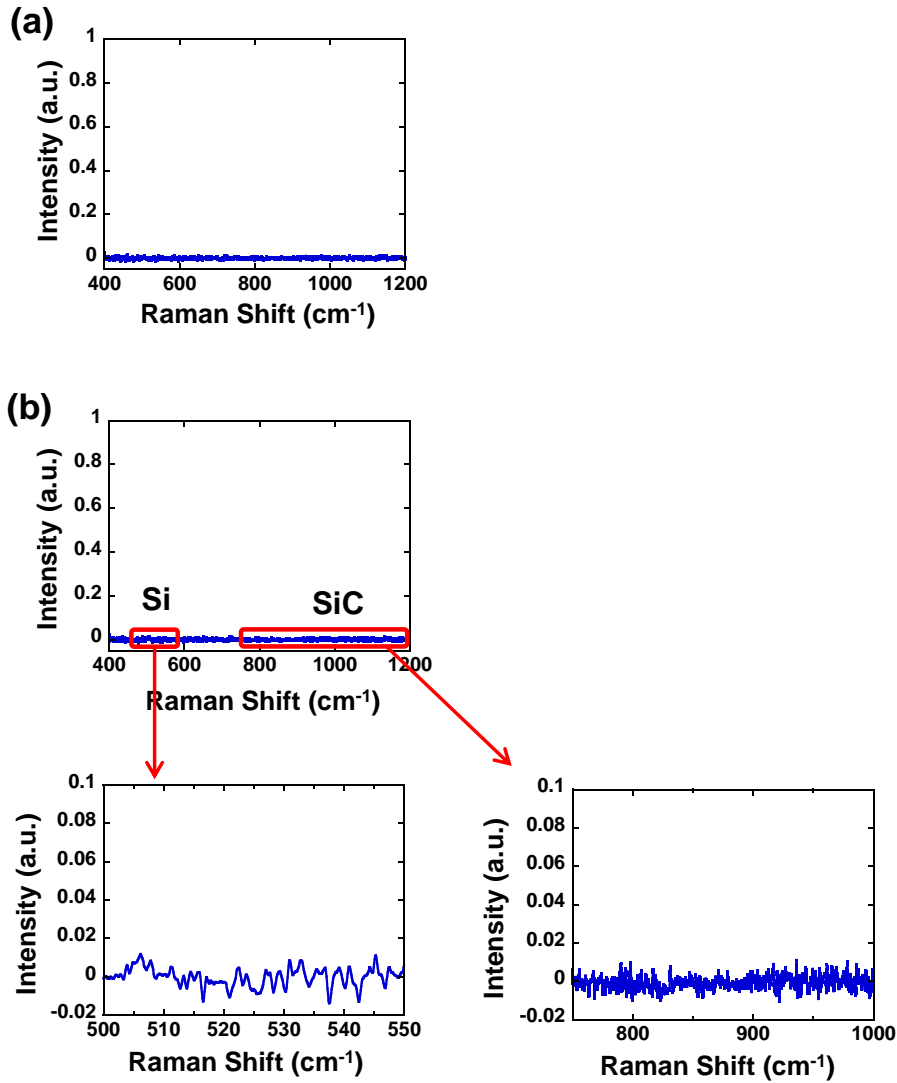

**Fig. S7.** Raman spectrum taken from **(a)** an undoped nSCD substrate and **(b)** a boron doped nSCD substrate. The bottom left and bottom right are the magnified views of the range of 500 ~ 550 cm$^{-1}$ and 700~1000 cm$^{-1}$, respectively.

## 8. I-V characteristics from the undoped and doped nSCD

Fig. S8 shows the R-V (Resistance vs. Voltage) characteristics by measuring two adjacent ohmic metals from the undoped (top) and the doped (down) nSCD. I−V characteristics were first measured and calculate the resistance to display the resistance between two samples. As shown in a Fig. S8, a boron doped nSCD via thermal diffusion doping has two orders lower resistance value. Also, the nearly linear I−V characteristics both from the undoped and doped nSCD suggest that a good ohmic contact were made.

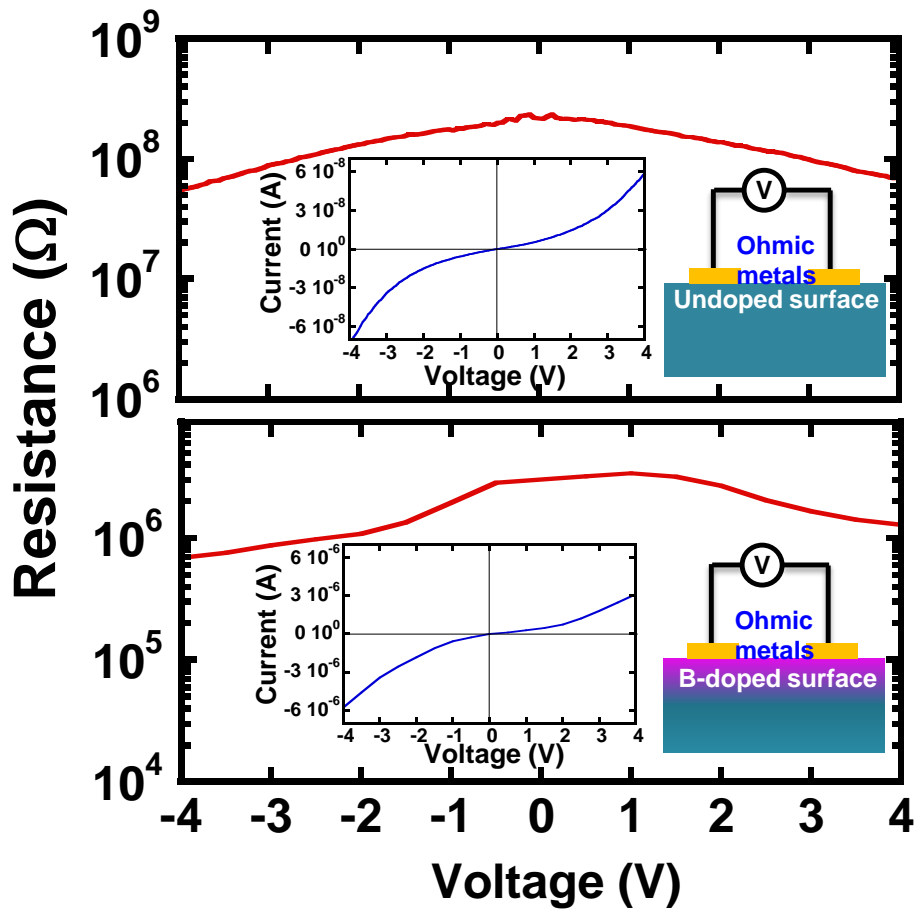

**Fig. S8.** An R−V (Resistance vs. Voltage) from the (top) undoped nSCD sample and (down) doped nSCD sample. Inset shows the measured I-V characteristics, respectively.

## 9. I-V equations for Schottky diode and for p-i diode

The current density of a Schottky diode is written as[3]

$$J_n = q \cdot v \cdot N_C \cdot \exp\left(-\frac{\phi_B}{V_t}\right) \cdot \left(\exp\left\{\frac{V_a}{V_t}\right\} - 1\right) \quad [1]$$

The current density of a Schottky diode is directly proportional to the doing concentration ($N_C$) of the semiconductor (drift) region, as shown in Eq. 1. The diamond diode reported by Brezeanu, M. et al.[4] is a Schottky diode. The reported doping concentration in the drift (i.e., intrinsic) region of the Schottky diode is as high as $5 \times 10^{15}$ cm$^{-3}$, which results in a high current density.

The current density of a p-n diode is written as (3)

$$J = J_N \cdot (-x_p) + J_P \cdot (x_n) = q\left(\frac{D_N}{L_N} \cdot \frac{n_i^2}{N_A} + \frac{D_P}{L_P} \cdot \frac{n_i^2}{N_D}\right) \cdot \left(\exp\left\{\frac{q \cdot V_a}{kT}\right\} - 1\right) \quad [2]$$

For a one-sided p+-n diode, the above equation can be re-written as

$$J = J_P \cdot (x_n) = q\left(\frac{D_P}{L_P} \cdot \frac{n_i^2}{N_D}\right) \cdot \left(\exp\left\{\frac{q \cdot V_a}{kT}\right\} - 1\right) \quad [3]$$

The diode demonstrated in this work is a *p-i* diode. The *p-i* diode is like a one-sided *p+-i* junction. The diode current density is related to the diamond intrinsic carrier concentration ($n_i$) and the doping concentration of the lightly doped side (i.e, the *i*-side), $N_D$. Since the i-region of our nSCD has no intentional doping, $N_D$ in theory is nearly equal to $n_i$. Considering the very low intrinsic carrier concentration in our nSCD, which is mainly from thermal activation and some possible contributions from unintentional impurities and thus much lower than $5 \times 10^{15}$ cm$^{-3}$, the current density of our diode is much lower than the Schottky diode published by Brezeanu, M. et al.[4].